# Expanding the Medical Decathlon dataset: segmentation of colon and colorectal cancer from computed tomography images

**I.M. Chernenkiy** ✉
Neural Network Technologies Center
Institute of Urology and Human Reproductive Health
I.M. Sechenov First Moscow State Medical University of the Ministry of Health of the Russian Federation (Sechenov University).
8, Trubetskaya str., building 2, Moscow, 119991, Russia
chernenkiy_i_m@staff.sechenov.ru

**Y.A. Drach**
Department of Biomedical Technologies and Systems
N.E. Bauman Moscow State Technical University
2nd Baumanskaya St., 5, Bldg. 1, Moscow, 105005, Russia
dyaa19l127@student.bmstu.ru

**S.R. Mustakimova**
Department of Higher Mathematics, Mechanics and Mathematical Modeling
I.M. Sechenov First Moscow State Medical University of the Ministry of Health of the Russian Federation (Sechenov University).
8, Trubetskaya str., building 2, Moscow, 119991, Russia
mustakimova_s_r@student.sechenov.ru

**V.V. Kazantseva**
Department of Higher Mathematics, Mechanics and Mathematical Modeling
I.M. Sechenov First Moscow State Medical University of the Ministry of Health of the Russian Federation (Sechenov University).
8, Trubetskaya str., building 2, Moscow, 119991, Russia
kazantseva_v_v@student.sechenov.ru

**N.A. Ushakov**
Department of Higher Mathematics, Mechanics and Mathematical Modeling
I.M. Sechenov First Moscow State Medical University of the Ministry of Health of the Russian Federation (Sechenov University).
8, Trubetskaya str., building 2, Moscow, 119991, Russia
ushakov_n_a1@mail.student.sechenov.ru

**S.K. Efetov**
Hospital of Faculty Surgery No. 2, UKB No. 4
I.M. Sechenov First Moscow State Medical University of the Ministry of Health of the Russian Federation (Sechenov University)
Dovatora str. 15. Moscow, 119991, Russia
efetov_s_k@staff.sechenov.ru

**M.V. Feldsherov**
University Clinical Hospital No. 2
I.M. Sechenov First Moscow State Medical University of the Ministry of Health of the Russian Federation (Sechenov University)
8, Trubetskaya str., building 2, Moscow, 119991, Russia
feldsherov_m_v@staff.sechenov.ru

## Abstract

Colorectal cancer is the third-most common cancer in the Western Hemisphere. The segmentation of colorectal and colorectal cancer by computed tomography is an urgent problem in medicine. Indeed, a system capable of solving this problem will enable the detection of colorectal cancer at early stages of the disease, facilitate the search for pathology by the radiologist, and significantly accelerate the process of diagnosing the disease. However, scientific publications on medical image processing mostly use closed, non-public data. This paper presents an extension of the Medical Decathlon dataset with colorectal markups in order to improve the quality of segmentation algorithms. An experienced radiologist validated the data, categorized it into subsets by quality, and published it in the public domain. Based on the obtained results, we trained neural network models of the UNet architecture with 5-part cross-validation and achieved a Dice metric quality of 0.6988 ± 0.3. The published markups will improve the quality of colorectal cancer detection and simplify the radiologist's job for study description.

**Keywords** *Colorectal cancer · Deep learning · nnU-Net · Computed tomography · Computer vision*

# 1 Introduction

Colorectal cancer (CRC) is the third most common cancer in the Western Hemisphere, and its incidence increases with age. Metastasis to regional lymph nodes is a defining characteristic of CRC. Up to 20% of patients have metastatic disease, with the most frequent localization of metastases in the liver [1]. In addition to population aging and dietary habits in high-income countries, unfavorable risk factors such as obesity, lack of physical activity, and smoking increase the risk of CRC. [2]. Prior to treatment, multislice computed tomography (MSCT) with intravenous contrast is employed to diagnose colorectal cancer and assess the stage of tumorigenesis. MSCT also enables the physician to conduct surgical planning. The degree of tumor process spreading is determined according to the TNM classification system, which categorizes tumors based on the number of layers they have grown into the intestinal wall (T), whether regional lymph nodes are affected by metastases (N), and whether the tumor process has spread to other organs and systems (M) [10]. Automatic segmentation of the colon and its neoplasms would improve the process of surgical planning.

# 2 Research review

Artificial intelligence (AI) applications have transformed numerous industries and are widely used in a variety of consumer products and services. In medicine, AI is mainly used for image classification and natural language processing, with the potential to impact image-related specialties such as radiology, pathology, and gastroenterology (GE) [3]. Currently, the diagnostic performance of colonoscopy remains unsatisfactory with unstable accuracy. The convolutional neural network system (CNNS) has demonstrated its potential to assist endoscopists in improving diagnostic accuracy. Colorectal cancer has a high incidence rate worldwide, yet early detection significantly improves survival rates. [4] However, there are a number of limitations to the CNN system and controversy over whether it provides better diagnostic performance compared to human endoscopists [2, 5].

For the automatic segmentation of radial imaging modalities, the UNet architecture is the most commonly utilized neural network system (CNNS) [11]. The authors of [8] collected 1131 marked 2D images of colon tumors with a resolution of $0.779 \times 0.779 \times 5 mm^3$ from 158 multispiral computed tomography (MSCT) studies of patients. The UNet model was trained on these 2D slices, resulting in a Dice metric of 0.7089.

To consider the spatial location of an object, a three-dimensional image can be employed as input data. For instance, the authors of the paper [7] proposed the use of the Deep Colorectal Coordinate Transform (DeepCRC). Given that the colon has a single pathway and continuous structure extending between the blind colon and rectum anatomically, the authors suggest that a regression task could be employed as an additional training task. Each voxel of the colon is assigned a number between zero and one, with zero representing the location coordinate of the rectum and one representing the location coordinate of the blind colon. This transformation enabled the researchers to achieve a Dice metric value of 0.862 for the colon and 0.646 for colorectal cancer.

Magnetic resonance imaging (MRI) may also be performed on a patient to detect colorectal cancer. The authors trained a neural network based on 450 MRI studies that segments colon cancer with a Dice metric of 0.55 [12].

The collection and verification of data for computer vision in medicine is a labor-intensive process. A comprehensive dataset with colorectal and colorectal cancer markups could not be identified in the public domain. One of the most widely used datasets is a collection of abdominal MSCT studies in NifTi format from the open-source medical dataset "Medical Decathlon" [6]. One of the tasks in this dataset involves the storage of 126 CT images with colorectal cancer markups. A larger dataset of 1204 CT images with markups of 104 anatomical structures is also available [13]. Of these, 842 images have fully labeled colorectal masks with a resolution of $1.5 \times 1.5 \times 1.5 mm^3$, but lack labeling of colorectal pathologic changes.

The objectives of this study are twofold: firstly, to create a dataset comprising colorectal and tumor markers, and secondly, to assess the efficacy of colorectal markers in the detection of colorectal cancer. Our contribution is to extend the Medical Decathlon dataset [6], which is available at the following link: https://gitflic.ru/project/sechenovntc/mde_colon_segmentation.

# 3 Materials and methods

## 3.1 Data labeling

A dataset of abdominal MSCT studies in NifTi format from the open-source medical data repository Medical Decathlon [6] was utilized. To expedite the partitioning process, the MONAILabel plugin of the MONAI framework within the 3D Slicer program was employed. A model from [13] was leveraged as a trained model for colon segmentation. All false positives were manually removed by the authors of the aforementioned paper. Subsequently, a radiologist with 15 years of experience conducted a validation process, wherein the boundaries of the colon markup were verified on each slice. The existing colorectal cancer markings in the dataset remained unaltered. Validation by a radiologist reduced the size of the validated dataset to 122 studies. In this case, the 122 studies were categorized into three subsets based on the quality of the data: The "good" subset comprises 100 studies, while the "bad" subset contains 17 cropped studies (in which the entire colon is not visible on the image). The "bad" subset comprises five studies. Two of these studies were of poor quality and could not identify the entire colon. Two further studies involved colon stomas following surgery, while one study involved a hernia. Four studies were excluded from the set due to broken images, with axial plane images duplicated or the slice order mixed.
All subsets are publicly available. However, in the future, we will only use the "good" subset for model training. The "bad" subset can be used in the future to expand the training sample, but it is not recommended to validate colon segmentation. Only the "good" subset is suitable for validation, as it has all segments of the colon highlighted in it.

## 3.2 Method

The UNet neural network architecture was employed in this study. It was compiled using the Medical Open Network for AI (MONAI) library function.

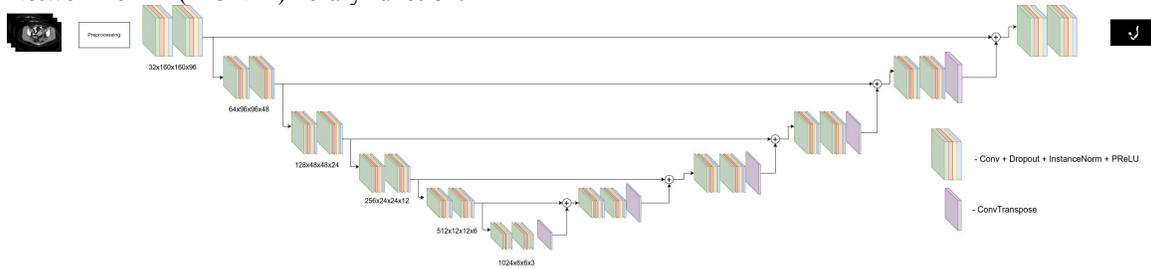

Figure 1 - Schematic diagram of the PlainUnet neural network architecture

As illustrated in Figure 1, the architecture is structured according to the principle of encoding and subsequent decoding of input data. Each level of encoder is associated with a specific set of convolutions, contingent upon the dimensions of the signal convolution kernel and the number of executed layer combinations.
The gray color indicates the principle of the residual signal, which is the original signal added to the signal transformed in the hidden layers of perceptron neurons.
The diagram also depicts a dropout layer, which randomly zeros out certain parts of the signal. This layer improves the generalization of the model by facilitating its transferability to data other than that on which it was trained.
In the context of medical data segmentation, where the objective is to identify a specific structure or region of interest within an image, the Dice loss function is employed to quantify the discrepancy between the predicted and true segmentation masks.
To enhance the quality, we employed the deep supervision approach [14]. Additionally, at each encoder level, we reduced the size of the original image mask to a smaller mask size. The proximity measure was performed with the reduced mask size to the required size at that level. Consequently, the loss function takes the form:

$$\mathcal{L}(\hat{y}, y) = \mathcal{L}_{DCE}(\widehat{y_0}, y_0) + \sum_{i=1}^{n} \mathcal{L}_{DCE}(\widehat{y_i}, y_i)$$

where $\mathcal{L}_{DCE}(x, y) = \mathcal{L}_D(x, y) + \mathcal{L}_{CE}(x, y)$ - where $\mathcal{L}_D$ – Dice loss function и $\mathcal{L}_{CE}$ is the cross-entropy loss function, $y_i$ – true segmentation mask for i-th layer, $\widehat{y_i}$ – is the predicted mask for the i-th layer.

In this study, a stochastic gradient descent (SGD) optimizer with Nesterov's impulse method [15], weight reduction of 3e-5, and a training step of 0.001 were utilized, with a reduction in the training step by a factor of 10 occurring at 625 epochs. The number of epochs was set to 1250. During the training phase, a batch size of two was generated with a cut region of size [160, 160, 96] voxels from a randomly selected location in the image with a resolution of 1.5x1.5x1.5x3 mm$^3$.

The PyTorch 1.13 framework and the MONAI 1.2.0 library were employed to train the model. The experiments were conducted on a machine with an NVIDIA Tesla V100 graphics card.

Prior to uploading the MSCT images to the trained neural network, data preprocessing was conducted by normalizing the images. This involved calculating the mean, standard deviation, and percentiles of 0.5 and 99.5 across the entire dataset, based on the voxels belonging to each class. Subsequently, the intersentile values were clipped within the percentile interval, followed by subtraction of the mean value and division by the standard deviation. In the training phase, the following augmentation techniques were employed to enhance the model's generalizability:

1) Mirroring. This operation mirrors one of the planes with a probability 0.5.
2) Athenic transformation. With a probability 0.2, the transformation rotates and scales a tensor.
3) Gaussian noise. With a probability 0.2, Gaussian noise is added to the tensor.
4) Sharpness change. With a probability of 0.2, the image is sharpened.
5) Intensity Scaling. With a probability of 0.15, the intensity of the tensor is modified within the range [0.75; 1.25].
6) Contrast change. With a probability of 0.1, the contrast of the image is altered by a coefficient in the range [0.7; 1.5].

For purposes of comparison with other methods, we undertook an analysis of the work of Isensee et al. [16], which we refer to as nnU-Net. This method has gained considerable popularity in the academic community for its application to medical image processing. In addition, we have access to pre-trained models on colorectal cancer segmentation from the Medical Decathlon dataset [6], which will allow us to assess the impact of adding a new class on segmentation quality.

A five-fold cross-validation procedure was employed on the "good" subset to obtain the results presented herein.

## 4 Results

The optimal mean value of the Dice metric was 0.6988 ± 0.3. Table 1 presents the results for each class. We attempted to utilize blocks with residual connections, yet they yielded inferior outcomes relative to the simple Unet with classical block Convolution+Normalization+Activation. Additionally, LayerNorm [17] and GroupNorm [16] were employed in place of InstanceNorm [18]. Figure 2 illustrates the partitioning of the EA and neural network on a single slice, along with the 3D construction of the markup.

Table 1. Dice metric results

|  | Mean value | Colon | Tumor |
| --- | --- | --- | --- |
| PlainUnet | 0,6254±0,3285 | 0,8082±0,1337 | 0,4425±0,3626 |
| nnU-Net (lowres) | 0,6988±0,3044 | 0,8651±0,0812 | 0,5324±0,3518 |

Additionally, the metrics were calculated when the "bad" subset was included in each model and across the entire dataset, excluding the colon, to determine the increase in the Dice metric with a model trained on tumor detection only. As the "bad" subset contains studies with not all colon segments, no filtering of large connected components was employed in the post-processing stage. When calculating the entire dataset, the original masks of the Medical Decathlon dataset were utilized to calculate the metrics [6]. Additionally, we disabled large component filtering in the post-processing stage. The results are presented in Table 2.

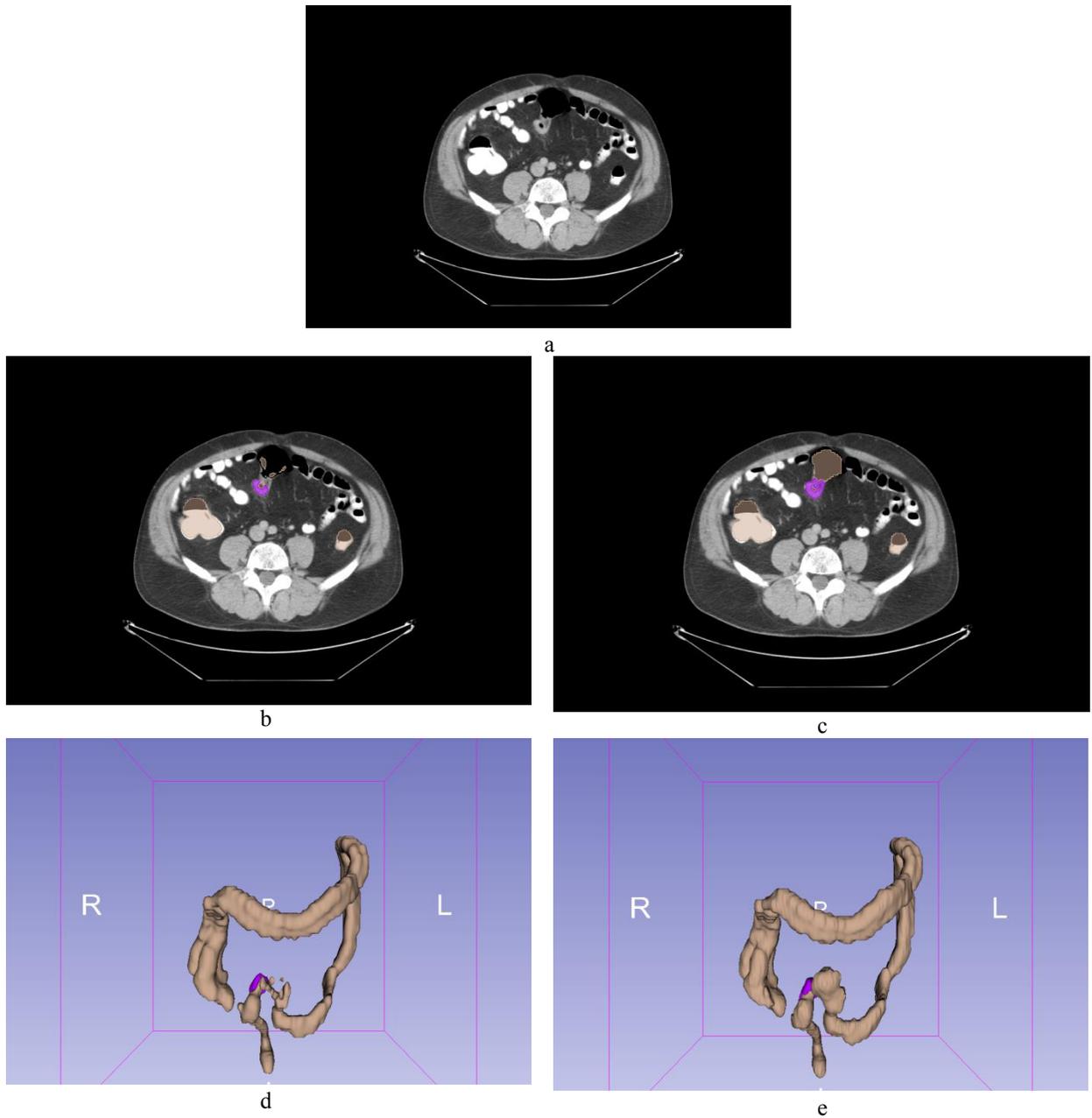

Figure 2 - The layout and 3D construction of a neural network (left) and an expert system (right). The figure includes axial slice (a), axial slice with superimposed masks of the neural network (b), axial slice with superimposed masks of expert (c), 3D constructions of masks of neural network (d) and 3D constructions of masks of expert (e).

Table 2. Metrics values for all sub-sets

| Model | Mean value | Colon | Tumor |
|---|---|---|---|
| PlainUNet + "bad" subset | 0,6218±0,3175 | 0,7971±0,1403 | 0,4464±0,3474 |
| nnU-Net lowres + "bad" subset | 0,6548±0,3225 | 0,8283±0,1296 | 0,4813±0,3624 |
| PlainUNet + entire dataset | - | - | 0,3891±0,3414 |
| nnU-Net lowres + entire dataset | - | - | 0,4294±0,3731 |

**Case review**

In this section, we aim to illustrate the efficacy of the proposed method through the presentation of a single case. We will provide metrics for both cases and formations. As metrics for demonstrating the case, we include metrics for comparing 3D regions. These include the Dice-Sorensen metric, the Hausdorff distance (which measures the largest distance between the boundaries of the true and predicted masks), and the average surface distance (which measures the average distance between the boundaries of the true and predicted masks).

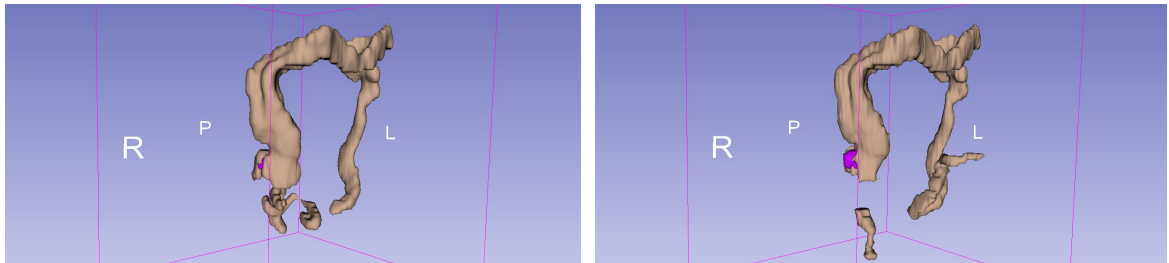

Figure 3. 3D plotting of two cases.

Figure 3 depicts the initial case from the "bad" subset in the top row. Its purpose is to illustrate the rationale behind the sorting of the case into a distinct set and to assess the efficacy of the model's performance on non-standard data. The patient had previously undergone resection of the sigmoid colon and had a stoma on the outside. The model result is presented on the left, and the expert's markup is presented on the right. The quality metrics yielded values for the colon of the Dice-Sorensen metric of 0.85, a Hausdorff distance of 57.31 px, and a mean surface distance of 3.38 px across all classes. In the case of the tumor, the following characteristics were obtained: Dice-Sorensen metric of 0.49, Hausdorff distance 8 px., and average surface distance of 0.78 px. It can be seen from this case that the model failed to detect the pathological course of the colon and falsely triggered the small intestine loops.

## 5 Conclusion

The Medical Decathlon dataset was augmented with colon markups based on publicly available data. Subsequently, experiments were conducted to train Unet-type architectures, resulting in an average Dice metric value of 0.6988±0.3. This represents the first public dataset with colon tumor and colon MSCT images. The marker data and code are available for download at the following link: https://gitflic.ru/project/sechenovntc/mde_colon_segmentation.

**Conflict of interests.** The authors declare that there is no conflict of interest.
**Financial support.** The study was not sponsored (it was funded by the authors' own resources).

**Acknowledgment.** The authors would like to express their gratitude to the Institute of Computer Science and Mathematical Modeling at Sechenov University for providing access to the computational cluster.